\documentclass{svproc}
\usepackage{url}

\usepackage{graphicx}
\usepackage{float}
\usepackage{multicol}
\usepackage{amsmath}
\usepackage{amssymb}
\def\sign{\mbox{sign\,}}

\begin{document}
\mainmatter              
\title{PLL and Costas loop based carrier recovery circuits for 4QAM:
non-linear analysis and simulation}
%
%

\author{N.V. Kuznetsov\inst{1}\inst{2}\inst{3} \and J.~Ladvanszky\inst{4} \and M.V. Yuldashev\inst{1} \and R.V. Yuldashev\inst{1}}
\authorrunning{N.V. Kuznetsov al.} 
%
%
\institute{Saint-Peterburg state University, Russia,\\
\email{nkuznetsov239@gmail.com},
\and
Institute of Problems of Mechanical Engineering RAS, Russia,
\and
Dept. of Mathematical Information Technology,University of Jyv\"{a}skyl\"{a}, Finland
\and 
Ericsson Telecom, Hungary}

\maketitle              

\begin{abstract}
Design of stable carrier recovery circuits
are used in many applications: wireless digital
communication, optical communication, microwave devices and other applications.
Quadrature Phase-Shift Keying (QPSK, 4-QAM) is used as modulation technique in many of these applications,
since QPSK provides double the data rate of classic Binary PSK (BPSK) modulation.
Analysis of Costas loop is a hard task because of its non-linearity.
In this work we consider two well-known modifications of 4QAM Costas loop circuits
and discuss a new circuit proposed by J.~Ladvanszky.
MATLAB Simulink models of the circuits are provided and the noise analysis is performed.
\keywords{PLL, non-linear analysis, hidden oscillations}
\end{abstract}

\section{Introduction}

Nowadays phase-locked loop based circuits are widely used in various intelligent systems for synchronization and communication \cite{intelligent-book,intelligent-book2,intelligent-book3}. Phase-shift keying (PSK) is a digital modulation process which conveys data by changing (modulating) the phase of a reference signal (the carrier wave).
The modulation occurs by varying the phase of sine and cosine inputs at a precise time. It is widely used for wireless networks (WiFi), radio-frequency identification (RFID),
and Bluetooth communication.
First widely adopted circuit for demodulation of Binary PSK signals (BPSK)
was invented by famous engineer John P. Costas \cite{Costas-1962-patent}.
Later it was adopted for more efficient modulation technique --- quadrature phase-shift keying
(also called QPSK, 4-PSK and 4-QAM) \cite{BestKLYY-2016,Best-2018}.
With four phases, QPSK can encode two bits per symbol,
while BPSK can encode only one bit per symbol.
More sophisticated modulation techniques (8-QAM, 16-QAM and higher)
require transmission with high signal-to-noise ratio.
One of the main characteristics of Costas loops is the lock-in range
(introduced in \cite{Gardner-1966},\cite[p.70]{Gardner-1979-book}).
Here we follow the rigorous mathematical definition of lock-in range for PLL-based circuits suggested in \cite{KuznetsovLYY-2015-IFAC-Ranges,LeonovKYY-2015-TCAS,BestKLYY-2016}.

In this work we study various models of QPSK Costas loop
including a new model based on the folding operation
(corresponding consideration of BPSK Costas loop can be found, e.g. in \cite{BestKLYY-2014-IJAC,LeonovKYY-2015-SIGPRO,BestKKLYY-2015-ACC,BestKLYY-2016,ladvanszky2018costas}).
Often only linear analysis and straight-forward circuit simulation is used
to study lock-in range, which may lead to misleading conclusions.
In this work non-linear mathematical models for each circuits are developed
and the lock-in ranges for these models are computed.
Finally, we compare noise characteristics of considered circuits (see, e.g. \cite{ladvanszky2011software}).

\subsection{Classical QPSK Costas loop}
\begin{figure}[H]
  \centering
  \includegraphics[width=0.5\linewidth]{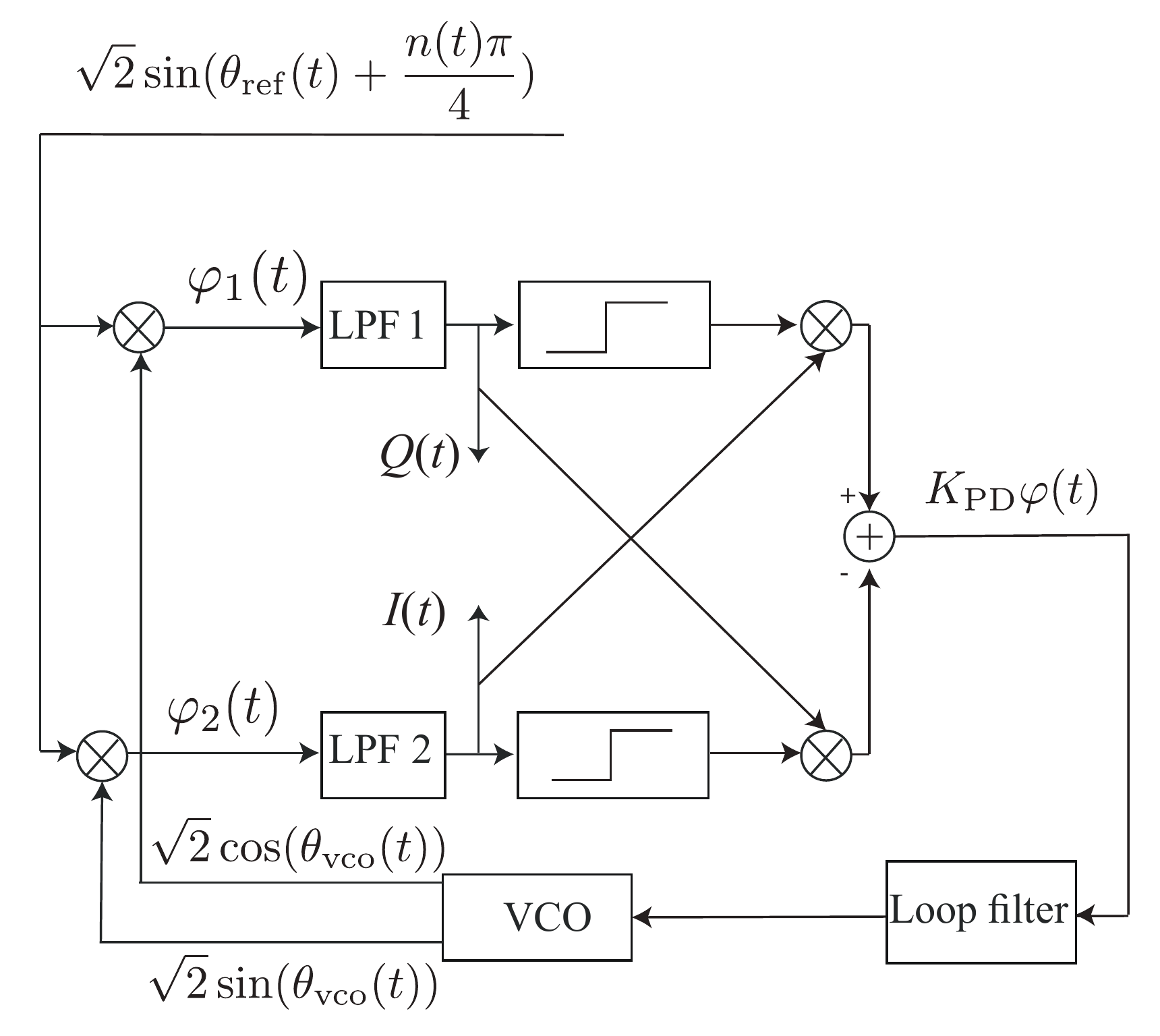}
  \caption{QPSK Costas loop operation.}
\label{fig:costas_after_sync}
\end{figure}

Consider the classical QPSK Costas loop operation (see Fig.~\ref{fig:costas_after_sync}).
It is convenient to consider input QPSK signal in the following form
\begin{equation}
\nonumber
\begin{aligned}
\sqrt{2}\sin(\theta_{\rm ref}(t)+\tfrac{n(t)\pi}{4}),
    \quad \theta_{\rm ref}(t) = \omega_{\rm ref}t,
    \ \  n(t) \in \{1,3,5,7\}.
\end{aligned}
\end{equation}
Here $\dot\theta_{\rm ref}(t) = \omega_{\rm ref}$ denotes carrier frequency and $n(t)$ corresponds to digital data (two bits per symbol).
The input signal is multiplied by inphase and quadrature phase VCO outputs $\sqrt{2}\cos(\theta_{\rm vco}(t))$ and $\sqrt{2}\sin(\theta_{\rm vco}(t))$, with $\theta_{\rm vco}(t)$ being the phase of VCO.
The resulting signals are
\begin{equation}
  \nonumber
  \begin{aligned}
      &
      \varphi_1(t) =
      2\cos(\theta_{\rm vco}(t))
      \sin(\theta_{\rm ref}(t)+\tfrac{n(t)\pi}{4}) =
      \\
      &
      =
      \sin(\theta_{\rm ref}(t)+\tfrac{n(t)\pi}{4} - \theta_{\rm vco}(t))
      +
      \\
      &
      \qquad+
      \sin(\theta_{\rm ref}(t)+\tfrac{n(t)\pi}{4} + \theta_{\rm vco}(t)),\\
      &
      \varphi_2(t) =
      2\sin(\theta_{\rm vco}(t))
      \sin(\theta_{\rm ref}(t)+\tfrac{n(t)\pi}{4})
      =
      \\
      &
      =
      \cos(\theta_{\rm ref}(t)+\tfrac{n(t)\pi}{4} - \theta_{\rm vco}(t))
      -
      \\
      &
      \qquad-
      \cos(\theta_{\rm ref}(t)+\tfrac{n(t)\pi}{4} + \theta_{\rm vco}(t))
  \end{aligned}
\end{equation}




Here, from an engineering point of view,
the high-frequency terms
$\cos(\theta_{\rm ref}(t)+ \theta_{\rm vco}(t) + \tfrac{n(t)\pi}{4} )$
 and $\sin(\theta_{\rm ref}(t) + \theta_{\rm vco}(t) + \tfrac{n(t)\pi}{4})$
are removed by low-pass filters LPF 1 and LPF 2\footnote{
While this is reasonable from a practical point of view,
its use in the analysis of Costas loop
requires further consideration
(see, e.g., \cite{PiqueiraM-2003}).
The application of averaging methods
allows one to justify the Assumption
and obtain the conditions under which it can be used
(see, e.g., \cite{LeonovKYY-2012-TCASII,LeonovKYY-2016-DAN}).
}.
Thus,
the signals $Q(t)$ and $I(t)$ on the upper and lower branches
can be approximated as
\begin{equation}\label{g1g2-approx}
  \begin{aligned}
    &
      Q(t) \approx
       \sin(\theta_{e}(t)+\tfrac{n(t)\pi}{4}),
    \\
    &
      I(t) \approx
      \cos(\theta_{e}(t)+\tfrac{n(t)\pi}{4}),
    \\
    &
      \theta_e(t) = \theta_{\rm ref}(t) - \theta_{\rm vco}(t).
  \end{aligned}
\end{equation}

Signals $I(t)$ and $Q(t)$ are used for demodulation, i.e. decoding $n(t)$.
In the case of complete synchronization of VCO to the carrier, i.e. $\theta_e(t) = 0$, one can determine $n(t)$ from signs of $I(t)$ and $Q(t)$ using the following table:
\begin{center}
    \begin{tabular}{ | l | l | l | p{5cm} |}
    \hline
    & $I>0$ & $I<0$ \\ \hline
    $Q>0$ & $n=1$ & $n=3$ \\ \hline
    $Q<0$ & $n=7$& $n=5$ \\ \hline
    \end{tabular}
\end{center}
\smallskip

\begin{figure}[H]
  \centering
  \includegraphics[scale=0.4]{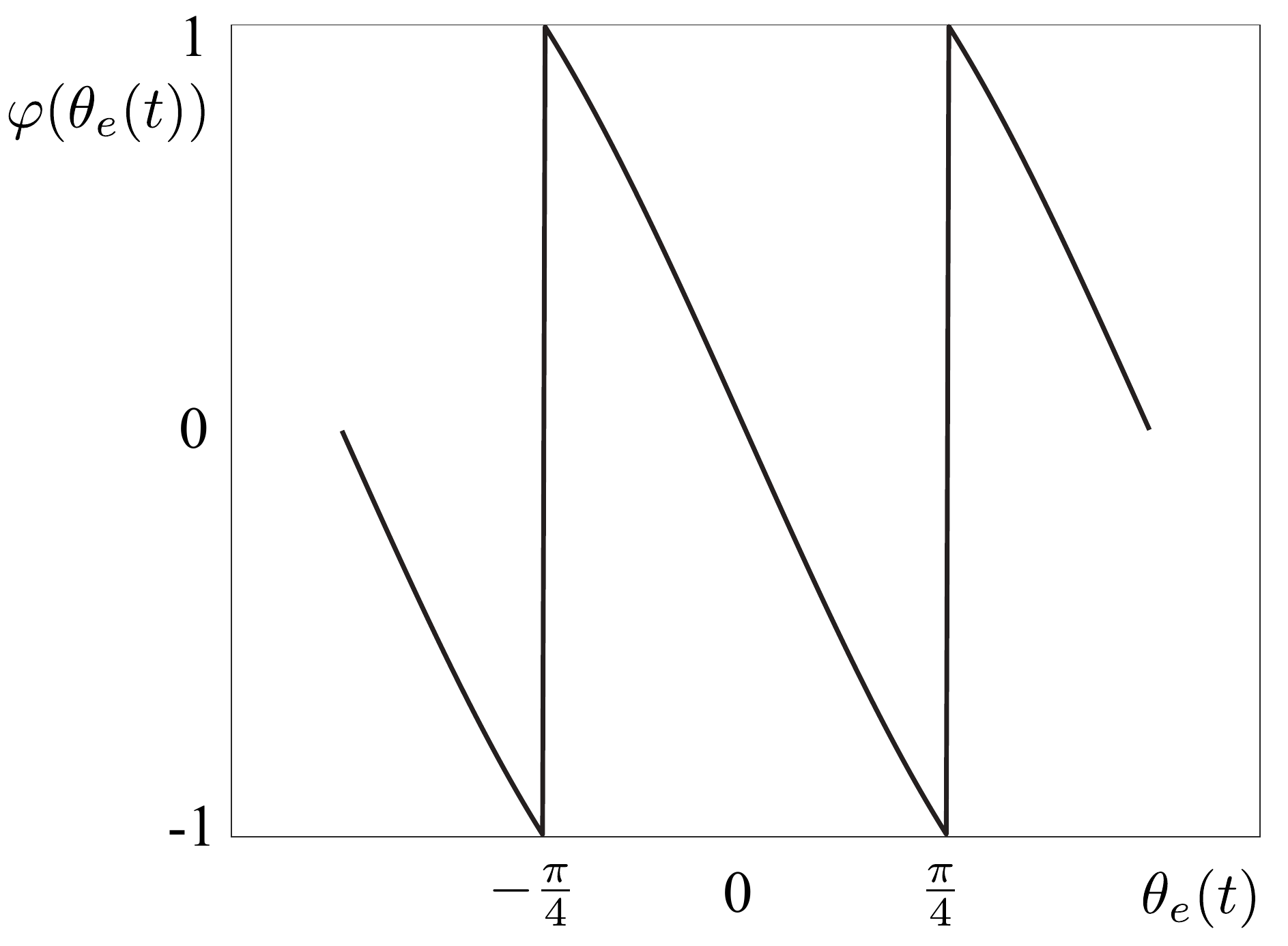}
\caption{Phase detector characteristic of QPSK Costas loop $\varphi(\theta_e)$.}
\label{fig:costas_out}
\end{figure}

After the filtration, both signals $\varphi_1(t)$ and $\varphi_2(t)$
pass through the limiters.
Then the outputs of the limiters $\sign\big(Q(t)\big)$ and $\sign\big(I(t)\big)$
are multiplied by $I(t)$ and $Q(t)$.
Using \eqref{g1g2-approx}
we get (see, e.g. \cite{BestKLYY-2016})
\begin{equation}\label{loop-filter-input-approx}
  \begin{aligned}
   &  K_{\rm PD}\varphi(t)
     = I(t)\sign\big(Q(t)\big) - Q(t)\sign\big(I(t)\big)\\
     & =\cos(\theta_{e}(t)+\tfrac{n(t)\pi}{4})
      \sign(\sin(\theta_{e}(t)+\tfrac{n(t)\pi}{4}))
      -
      \\
      &
      -
      \sin(\theta_{e}(t)+\tfrac{n(t)\pi}{4})
      \sign(\cos(\theta_{e}(t)+\tfrac{n(t)\pi}{4}))
    \\
    &
    =\,K_{\rm PD}\varphi(\theta_e(t))\,=\,
    \left\{
      \begin{array}{ll}
        -\frac{1}{\sqrt{2}}\sin(\theta_e(t)), &  -\frac{\pi}{4}< \theta_e(t) < \frac{\pi}{4}, \\
        \frac{1}{\sqrt{2}}\cos(\theta_e(t)), &  \frac{\pi}{4}< \theta_e(t) < \frac{3\pi}{4}, \\
        \frac{1}{\sqrt{2}}\sin(\theta_e(t)), &  \frac{3\pi}{4}< \theta_e(t) < \frac{5\pi}{4}, \\
        -\frac{1}{\sqrt{2}}\cos(\theta_e(t)), &  \frac{5\pi}{4}< \theta_e(t) < -\frac{\pi}{4}. \\
      \end{array}
   \right.
  \end{aligned}
\end{equation}
Here $K_{\rm PD} = \frac{1}{2}$, $\varphi(\theta_e(t))$ is a piecewise-smooth $\frac{\pi}{2}$---periodic function (see Fig.~\ref{fig:costas_out}) with unit amplitude.
The periodicity of function $\varphi(\theta_e)$ makes the loop filter input $\varphi(t)$ insensitive to change of the data signal $n(t)$.
The resulting signal $\varphi(t)$
after the filtration by the loop filter forms the
control signal $g(t)$  for the VCO.

The relation between the input $\varphi(t)$
and the output $g(t)$ of the Loop filter has the form
\begin{equation}\label{loop-filter}
 \begin{aligned}
 & \frac{dx}{dt} = A x + b  K_{\rm PD}\varphi(t),
 \ g(t) = c^*x + h K_{\rm PD}\varphi(t).
 \end{aligned}
\end{equation}
Here $A$ is a constant matrix,
vector $x(t)$ is a filter state,
$b,c$ are constant vectors.
Corresponding transfer function takes the form
\begin{equation}
  H(s) = -c^{*}(A - sI)^{-1}b + h.
\end{equation}

The control signal $g(t)$ is used to adjust VCO frequency to the frequency of input carrier signal
\begin{equation} \label{vco first}
   \dot\theta_{\rm vco}(t) = \omega_{\rm vco}(t) = \omega_{\rm vco}^{\text{free}} + K_{\rm vco}g(t).
\end{equation}
Here $\omega_{\rm vco}^{\text{free}}$ is free-running frequency of VCO
and $K_{\rm vco}$ is VCO gain.

For the constant frequency of input carrier
\begin{equation}\label{omega1-const}
   \dot\theta_{\rm ref}(t) = \omega_{\rm ref}(t) \equiv \omega_{\rm ref},
\end{equation}
 equations \eqref{loop-filter-input-approx}-\eqref{vco first}
give the following mathematical model of Costas loop
\begin{equation} \label{mathmodel-class}
 \begin{aligned}
   & \frac{dx}{dt} = A x + bK_{\rm PD}\varphi(\theta_e),
   \\
   & \dot\theta_e =
   \omega_{\rm ref} - \omega_{\rm vco}^{\rm free}
    -K_{\rm vco}
   \Big(
    c^*x + hK_{\rm PD}\varphi(\theta_e)
   \Big). \\
 \end{aligned}
\end{equation}

\begin{figure}[H]
  \centering
  \includegraphics[scale=0.6]{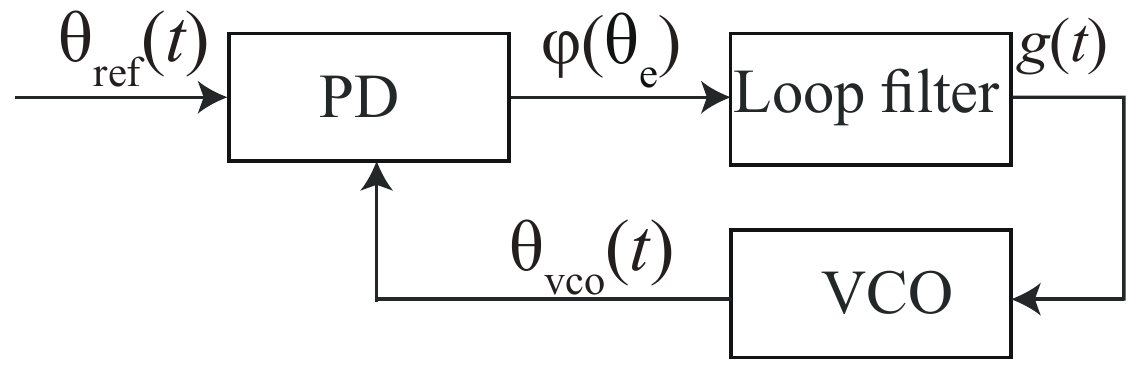}
  \caption{Classical mathematical model of QPSK Costas loop.}
  \label{fig:pll-qpsk}
\end{figure}

Non-linear mathematical model \eqref{mathmodel-class} and corresponding Fig.~\ref{fig:pll-qpsk}
is called \emph{the signal's phase space model} \cite{LeonovKYY-2015-TCAS}.
Engineers use this model for the approximate analysis of the circuit (see, e.g. \cite{Best-2018}).
Using the above signal's phase space model, further we estimate 
the lock-in ranges for the considered QPSK circuits.

\subsection{Fourth-power Costas loop}
\begin{figure}[H]
  \centering
  \includegraphics[width=\linewidth]{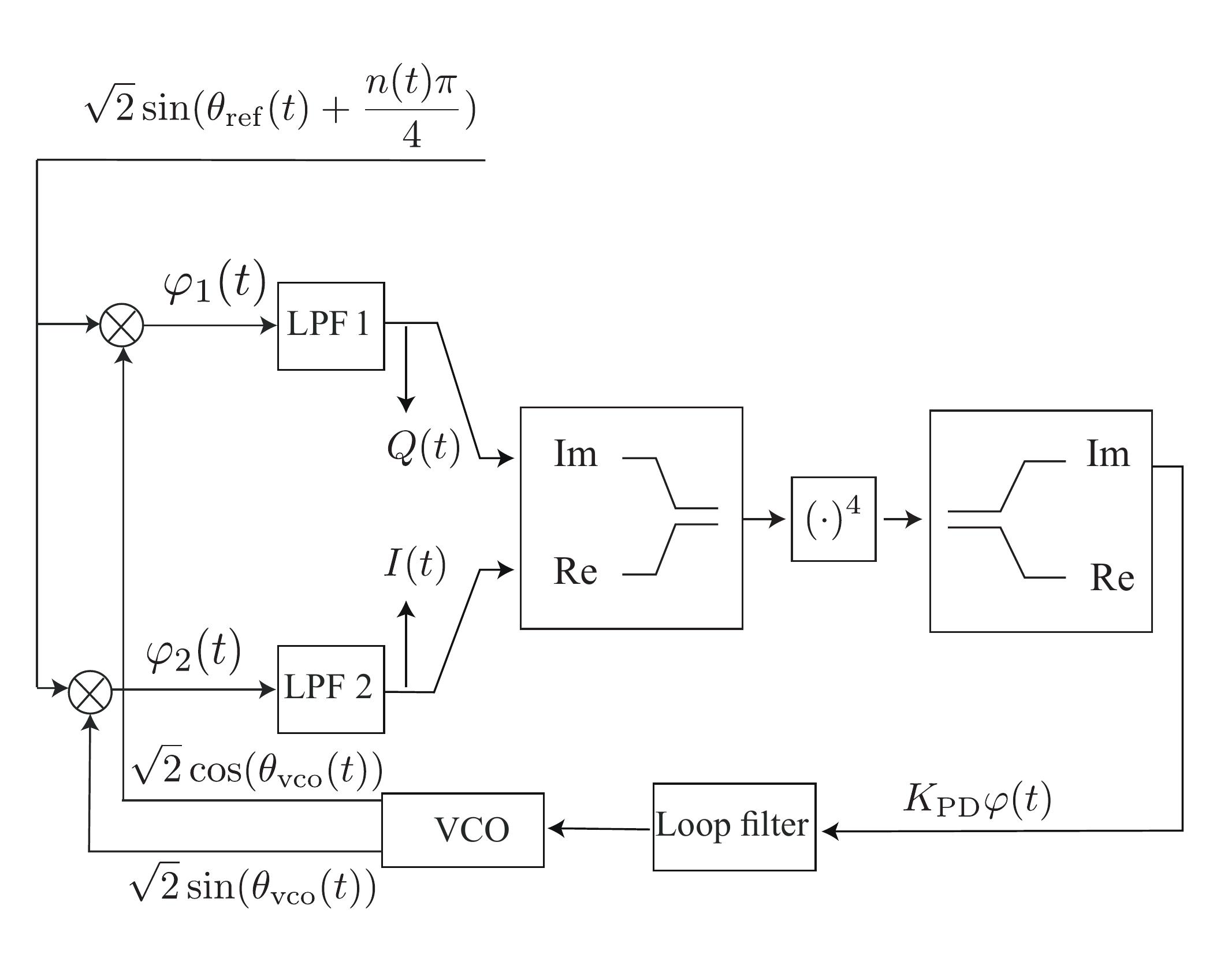}
  \caption{QPSK Costas loop with $(\cdot)^4$ block.}
  \label{fig:qpsk-janos-4th}
\end{figure}

Consider variation of QPSK Costas loop shown in Fig.~\ref{fig:qpsk-janos-4th} (see, e.g. \cite{weber1976candidate,osborne1982generalized}, and for optical implementations \cite{barry1992carrier}).
Similar to classic QPSK Costas loop (Fig.~\ref{fig:costas_after_sync}),
input QPSK signal is multiplied by two outputs of VCO on each branch and
outputs of multipliers filtered by corresponding low-pass filters.
Then signals from two branches ($I(t)$ and $Q(t)$) are combined by {\bf Re Im} block
to complex signal, which is raised to the fourth power,
and resulting signal is connected to input of VCO.
Using apporximation \eqref{g1g2-approx}, input of the loop filter can be approximated as follows
\begin{equation}
  \begin{aligned}
    &
      \varphi(\theta_e(t)) = \text{Re}(\left(I(t) + jQ(t)\right)^4
      \approx
        -\sin(4\theta_{e}(t))
  \end{aligned}
\end{equation}
Here phase detector characteristic $\varphi(\theta_e) = -\sin(4\theta_{e})$ is of unit amplitude.
Therefore we get the following equation
\begin{equation}
 \label{qpsk-4th-order-eq}
 \begin{aligned}
 & \dot{x} = A x - b \sin(4\theta_e),
 \\
 & \dot\theta_e =
  \omega_{e}^{\rm free}
  - K_{\rm vco}c^*x
  + K_{\rm vco}h\sin(4\theta_e).
 \end{aligned}
\end{equation}

\subsection{Folding QPSK Costas loop}
\begin{figure}[H]
  \centering
  \includegraphics[width=\linewidth]{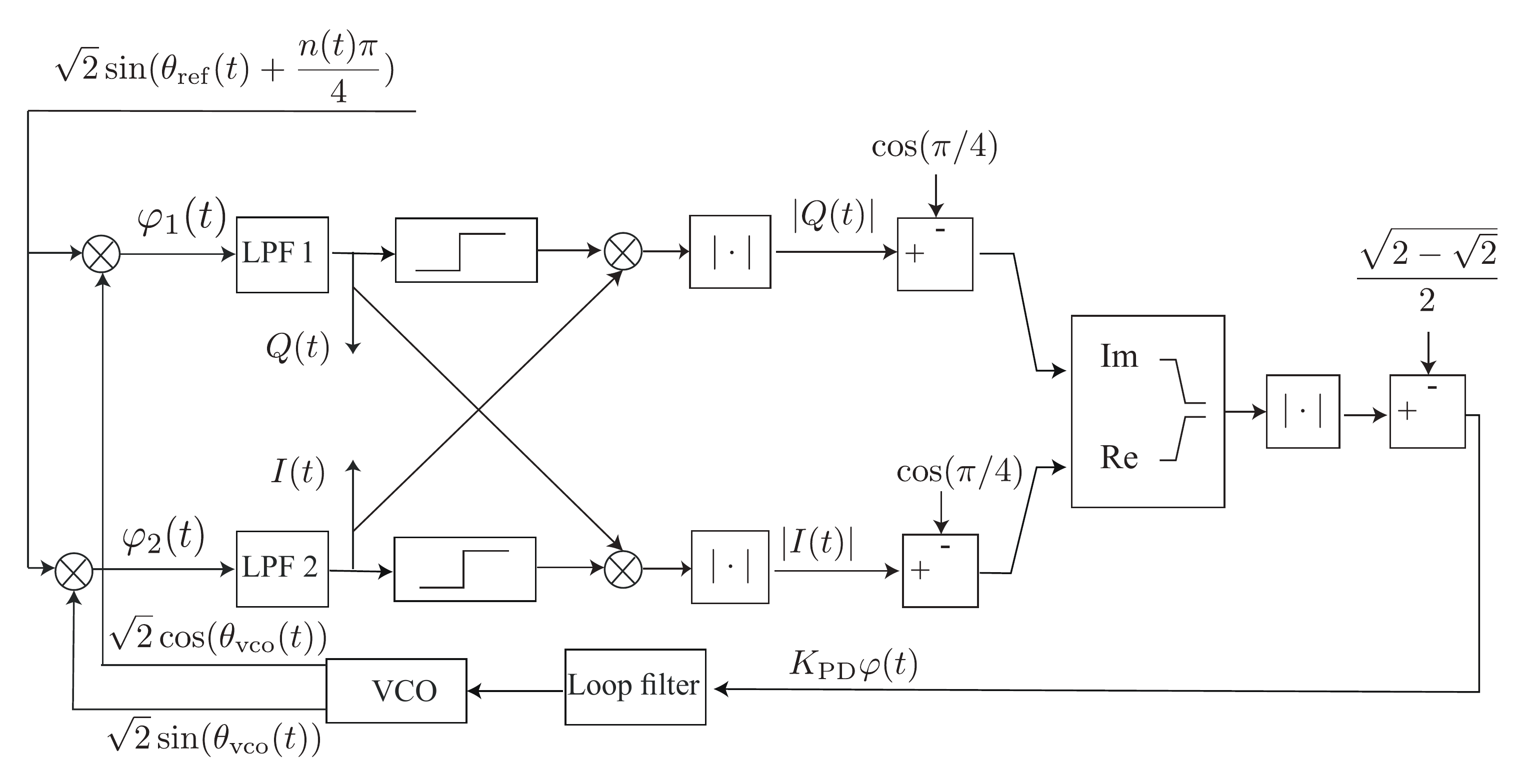}
  \caption{QPSK Costas loop with folding block.}
\label{fig:qpsk-folding-original}
\end{figure}

\begin{figure}[!htp]
  \centering
  \includegraphics[width=\linewidth]{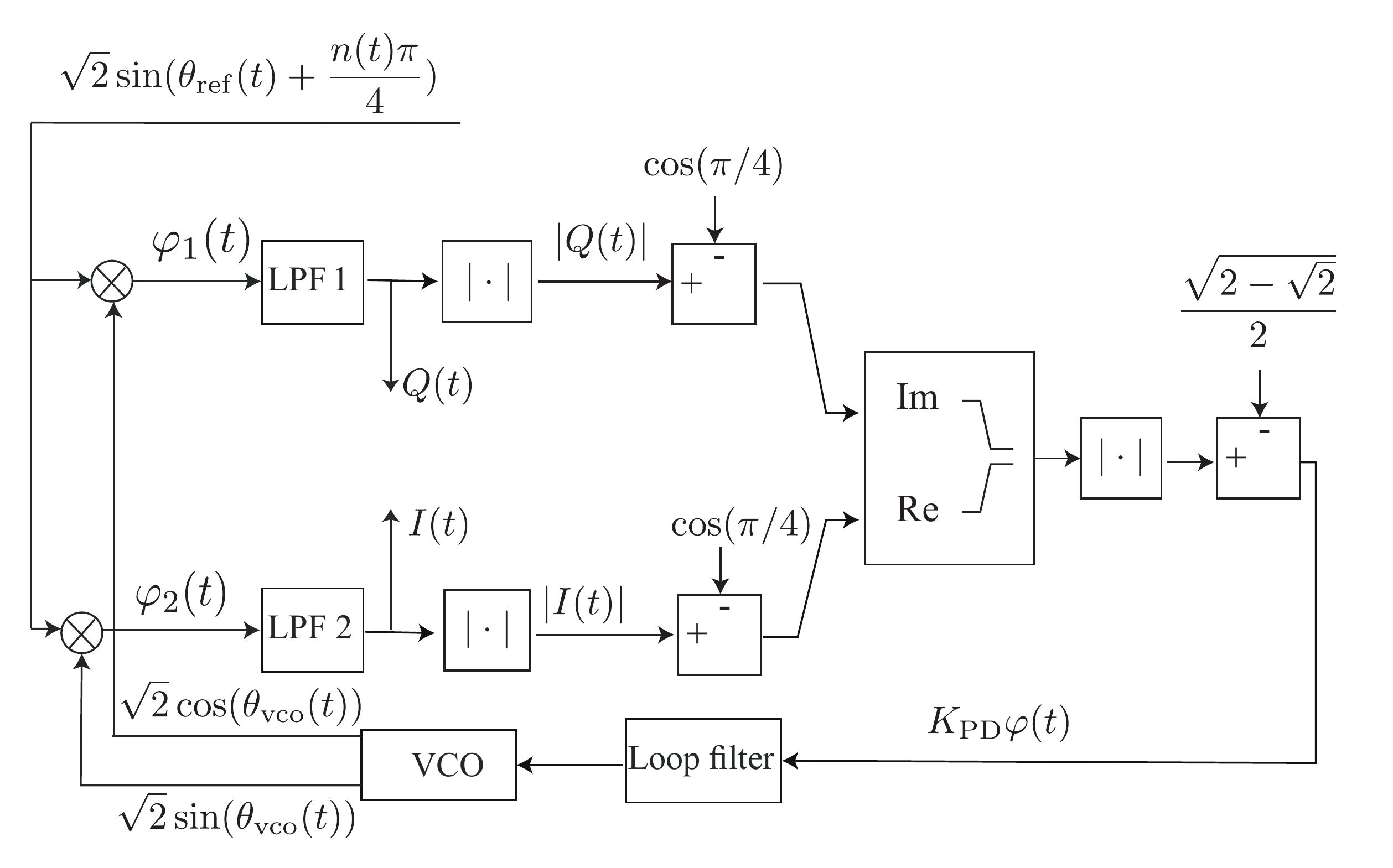}
  \caption{Simplified QPSK Costas loop with folding block.}
\label{fig:qpsk-simplified-folding}
\end{figure}

Now lets consider QPSK Costas loop proposed by J\'anos Ladv\'anszky
\cite{janos-2018-patent}
(see Fig.~\ref{fig:qpsk-folding-original}).
Here input signal, VCO output, multipliers and filters (LPF 1, LPF 2, Loop filter) are the same as for classic QPSK Costas loop.
As mentioned before, here we assume that low-pass filters completely eliminate the high-frequency oscillations.
Using the assumption of low-pass filtration, 
this scheme can be simplified, see  Fig.~\ref{fig:qpsk-simplified-folding}).
These two block-schemes are equivalent because blocks which take absolute values have the same outputs:
\begin{equation}
  \begin{aligned}
    & |\sin(\theta_e(t)+\frac{n(t)\pi}{4})\sign \cos(\theta_e(t)+\frac{n(t)\pi}{4})| =
    \\
    &
    = |\sin(\theta_e(t)+\frac{n(t)\pi}{4})|=|Q(t)|,\\
    & |\cos(\theta_e(t)+\frac{n(t)\pi}{4})\sign \sin(\theta_e(t)+\frac{n(t)\pi}{4})| =
    \\
    &
    = |\cos(\theta_e(t)+\frac{n(t)\pi}{4})|=|I(t)|,\\
  \end{aligned}
\end{equation}
These signals are shifted by $-\cos(\frac{\pi}{4})$ and combined by {\bf Re Im} block to complex signal
\begin{equation}
\begin{aligned}
  & |\cos(\theta_e(t)+\tfrac{\pi}{4})|-\cos(\tfrac{\pi}{4})
    +j(|\sin(\theta_e(t)+\tfrac{\pi}{4})|-\sin(\tfrac{\pi}{4})).
  \\
\end{aligned}
\end{equation}
Then absolute value of complex signal is centered,
and after filtration by Loop filter used as control signal for VCO:
\begin{equation}
\label{folding-characteristic}
  \begin{aligned}
    &
    K_{\rm PD}\varphi(\theta_e(t))
    = \big|
    |\cos(\theta_e(t)+\tfrac{\pi}{4})|-\cos(\tfrac{\pi}{4})
    +
    \\
    &
    +j(|\sin(\theta_e(t)+\tfrac{\pi}{4})|-\sin(\tfrac{\pi}{4}))
    \big|
    -\frac{\sqrt{2-\sqrt{2}}}{2}=\\
    & \sqrt{2-\sqrt{2}(|\sin(\theta_e(t)+\tfrac{\pi}{4})| + |\cos(\theta_e(t)+\tfrac{\pi}{4})|)}
    -\frac{\sqrt{2-\sqrt{2}}}{2}.
  \end{aligned}
\end{equation}
Here amplitude of \eqref{folding-characteristic} is $K_{\rm pd} = \frac{\sqrt{2-\sqrt{2}}}{2}$ and $\varphi(\theta_e)$ is normalized phase detector characteristic with unit amplitude\footnote{Alternatively Atomatic Gain Controll (AGC) circuits can be used to keep desired signal centered.
AGC allow to make circuit less sensitive to variations of input signal amplitudes,
however mathematical model will be the same.
Therefore here we consider only core model without engineering optimisations.
}.

\begin{figure}[H]
  \centering
  \includegraphics[width=0.4\linewidth]{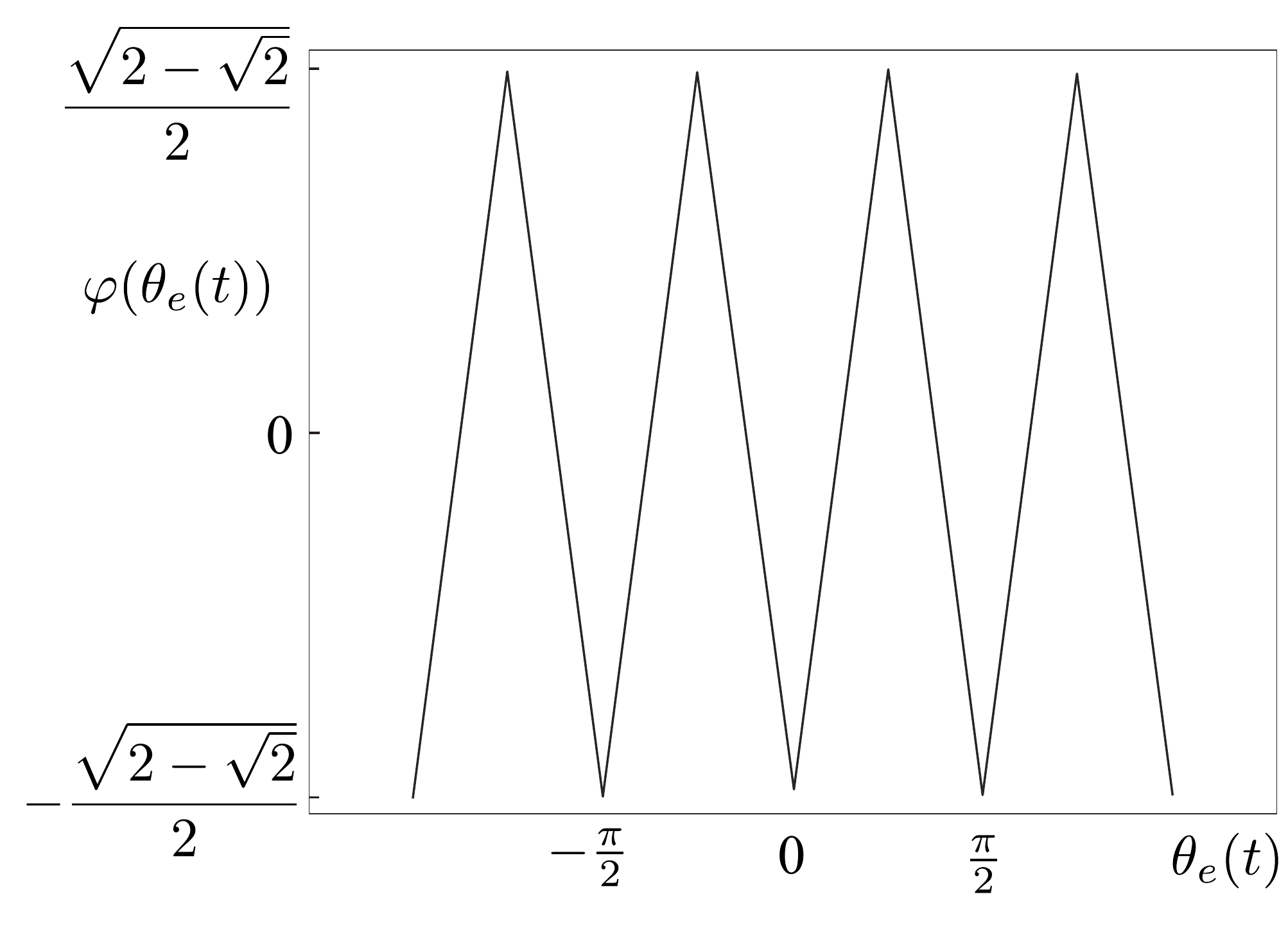}
  \caption{Phase detector characteristic for the folding QPSK Costas loop}
  \label{fig:folding-charactersitic}
\end{figure}
Note, that the folding version can work for higher order constellations as well,
however, the 4th power version can work only for 4QAM.

\section{Hold-in and lock-in ranges}
Since block-scheme on Fig.~\ref{fig:pll-qpsk} and corresponding equations \eqref{mathmodel-class} describe classical analog PLL, it is possible to apply the same analysis to QPSK Costas loops.
For proportionally-integrating loop filter the hold-in and pull-in ranges are infinite
\cite{LeonovKYY-2015-TCAS}.
The lock-in range can be recalculated from the lock-in range for the PLL with corresponding phase detector characteristics.

\begin{figure}[H]
  \centering
  \includegraphics[width=0.4\linewidth]{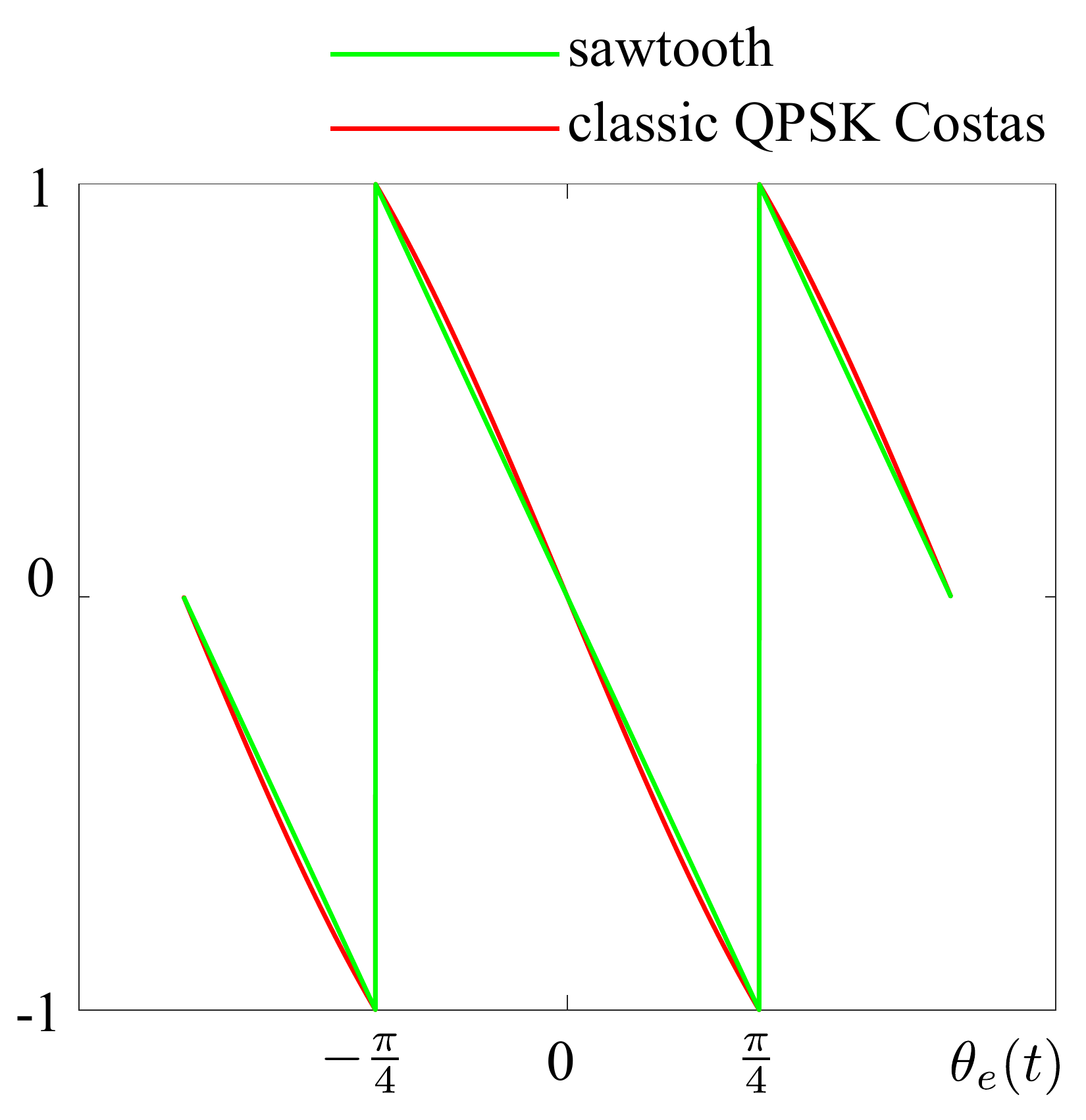}
  \caption{Normalized PD characteristic of the classic QPSK Costas loop
  and sawtooth characteristic}
  \label{fig:classic-vs-sawtooth}
\end{figure}

\begin{figure}[H]
  \centering
  \includegraphics[width=0.4\linewidth]{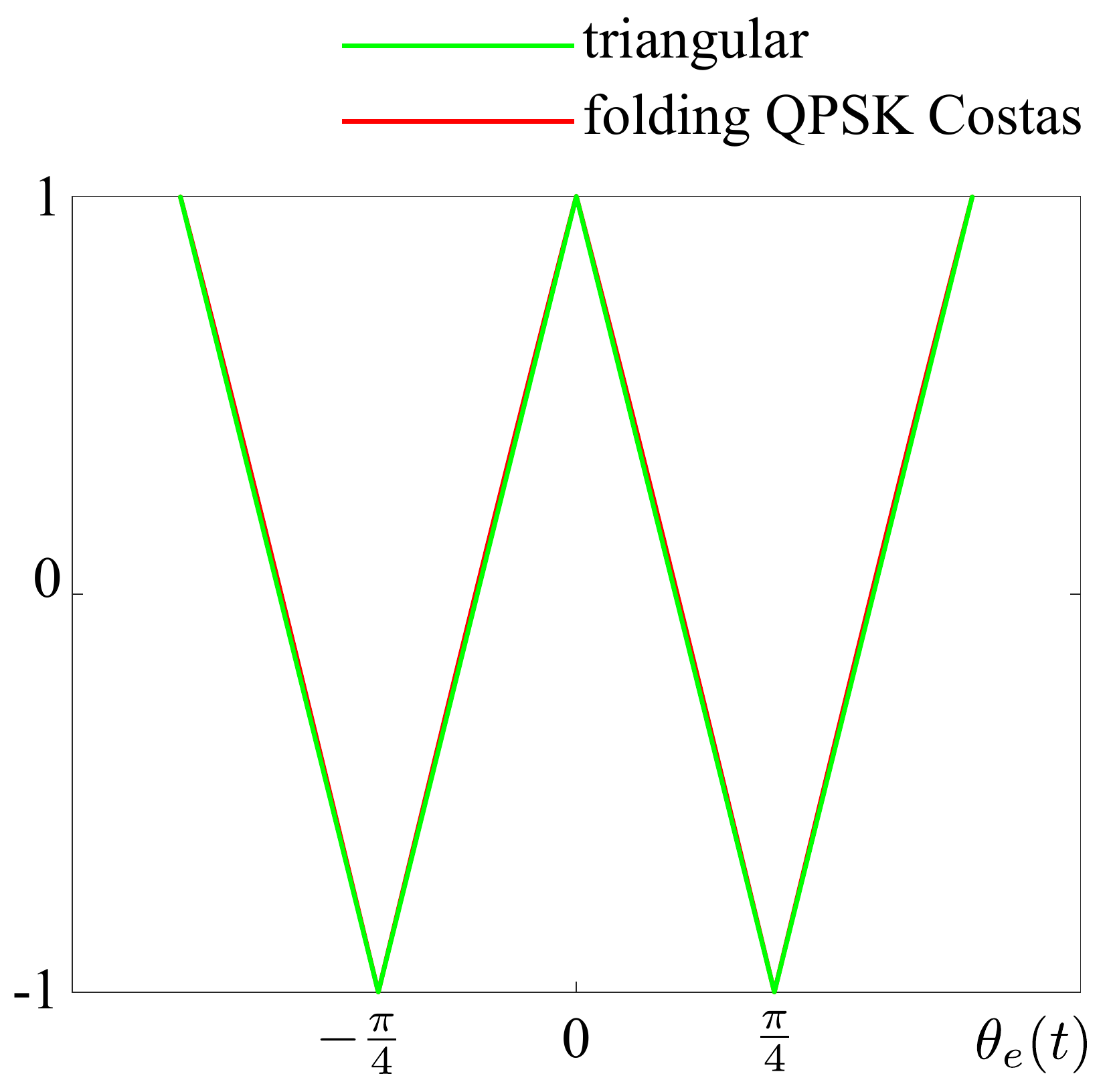}
  \caption{Normalized PD characteristic of folding QPSK Costas
  and triangular characteristic}
  \label{fig:folding-vs-triangular}
\end{figure}

Since PLL system \eqref{mathmodel-class} is $2\pi$-periodic (while for QPSK Costas it is $\frac{\pi}{2}$-periodic),
in order to transfer results to QPSK Costas loops it is necessary to consider  change of  variables
$\tilde\theta_e = 4\theta_e$, $\tilde \omega_e = \frac{\omega_{\rm ref} - \omega_{\rm vco}^{\rm free}}{4}$, $\tilde v_e(\tilde\theta_e)=\varphi(\theta_e)$.
After the change of variables we get
\begin{equation}
\label{phase-space qpsk}
 \begin{aligned}
   & \frac{dx}{dt} = A x - b\tilde v (\tilde\theta_e),
   \\
   & \dot{\tilde{\theta_e}}=
   4\omega_{\rm ref} - 4\omega_{\rm vco}^{\rm free}
    -4K_{\rm vco}
   \Big(
    c^*x - h\tilde v (\tilde\theta_e)
   \Big). \\
 \end{aligned}
\end{equation}
For sinusoidal characteristics (in the case of 4th power QPSK and
two-phase Costas suggested by Roland Best \cite{BestKLYY-2016})
$\varphi(\theta_e) = -\sin(4\theta_e)$ equations \eqref{phase-space qpsk} take the following form
\begin{equation}
\label{sinusoidal pll}
 \begin{aligned}
   & \frac{dx}{dt} = A x - b\sin(\tilde\theta_e),
   \\
   & \dot{\tilde{\theta_e}}=
   4\omega_{\rm ref} - 4\omega_{\rm vco}^{\rm free}
    -4K_{\rm vco}
   \Big(
    c^*x - h\sin(\tilde\theta_e)
   \Big). \\
 \end{aligned}
\end{equation}
Lock-in range for PLL-based circuits which are described by \eqref{sinusoidal pll}
is computed numerically and given in \cite{KuznetsovLYY-arxiv2017}.
For classical loop the phase-detector characteristic is almost sawtooth 
(maximum deviation is less than $0.05$, see Fig.~\ref{fig:classic-vs-sawtooth}).
For folding loop PD characteristics is almost triangular (maximum deviation is less than $0.03$, see Fig.~\ref{fig:folding-vs-triangular}).
Therefore we can use estimates of the lock-in range of classic PLL
with corresponding PD characteristics
\cite{LeonovKYY-2015-TCAS,AleksandrovKLYY-2016-arXiv-sin,AleksandrovKLYY-2016-arXiv-impulse}.

\subsection{Classical QPSK Costas}
\begin{align}
& \omega_l = 2\frac{a \sqrt{\pi}}{\tau_2} \hskip0.1cm \operatorname{exp}\left(\displaystyle \frac{a}{2 d_{-}} \ln \left( \frac{d_{+} + d_{-}}{d_{+} - d_{-}}\right)\right) ; \label{formula:LockInNode} \\
& \text{ where } a = \displaystyle \sqrt{\frac{4K_{\rm vco} \tau_2^2}{\tau_1}}, \hskip0.5cm d_{-} = |a|, \hskip0.5cm d_{+} = \sqrt{a^2 + 4\pi}. \nonumber
\end{align}
\subsection{4th power QPSK Costas}
\begin{figure}[H]
  \centering
  \includegraphics[width=\linewidth]{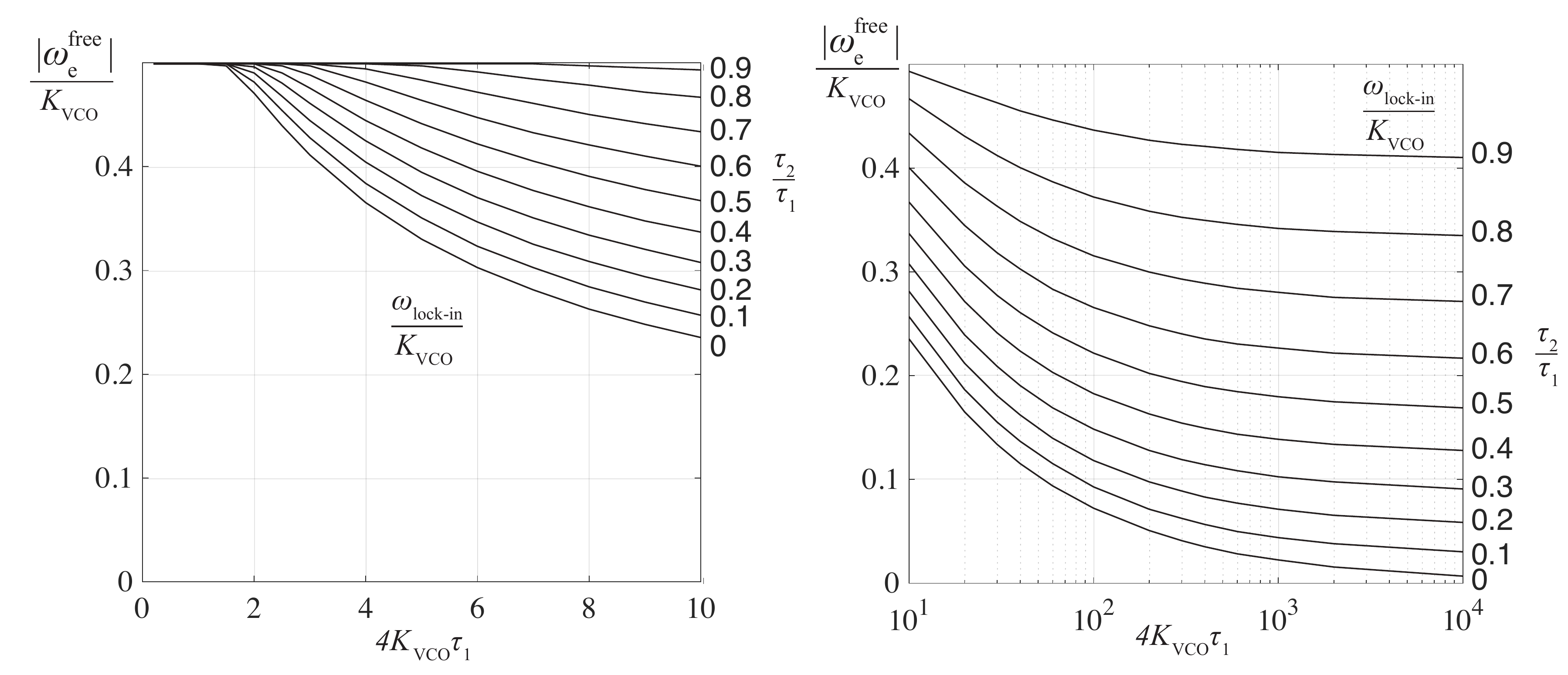}
  \caption{Lock-in range for 4th power QPSK Costas loop}
  \label{fig:4th-power-lockin}
\end{figure}
\subsection{Folding QPSK Costas} 
\begin{equation}
\begin{aligned}
& \textit{(i)} \text{ if } a^2 - 2\pi > 0,
  \\
  & \quad
  \text{ then } \omega_l = 2\frac{a \sqrt{\pi}}{\tau_2} \hskip0.1cm \operatorname{exp}\left(\displaystyle \frac{a}{2 d_{-}} \ln \left( \frac{d_{+} + d_{-}}{d_{+} - d_{-}}\right)\right) ;  \\
& \textit{(ii)} \text{ if } a^2 - 2\pi = 0,
  \\
  & \quad
  \text{ then } \omega_l = 2\frac{a\sqrt{\pi}}{\tau_2} \hskip0.1cm \operatorname{exp} \left(\displaystyle \frac{a}{d_{+}}\right);  \\
& \textit{(iii)} \text{ if } a^2 - 2\pi < 0,
  \\
  & \quad
  \text{ then } \omega_l = 2\frac{a\sqrt{\pi}}{\tau_2} \operatorname{exp}\left(\displaystyle \frac{a}{d_{-}} \operatorname{arctg}\left(\displaystyle \frac{d_{-}}{d_{+}} \right)\right),  \\[1.1em]
& \text{ where } a = \displaystyle \sqrt{\frac{4K_{\rm vco}K_{\rm pd} \tau_2^2}{\tau_1}},
\\
& \qquad\quad
 d_{-} = \sqrt{\Big|a^2 - 2\pi\Big|},
\\
& \qquad\quad
 d_{+} = \sqrt{|a^2 - 4\pi|}. \nonumber
\end{aligned}
\end{equation}

\section{Simulink implementations}
\begin{figure}[H]
  \centering
  \includegraphics[width=\linewidth]{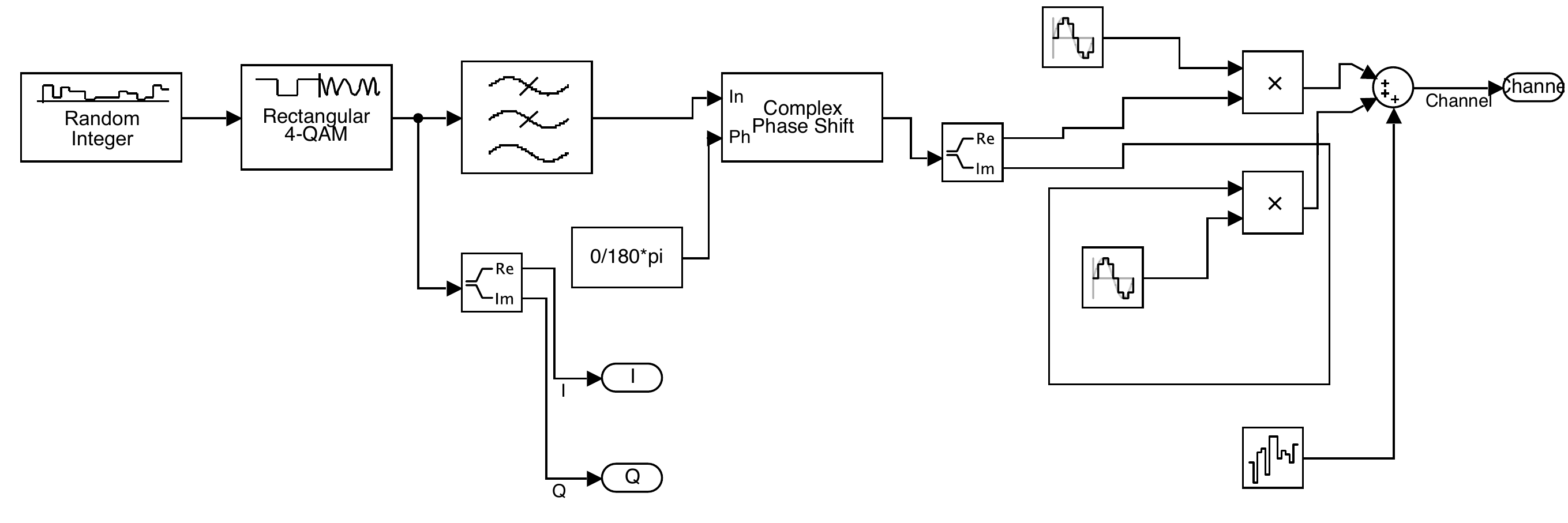}
  \caption{QPSK signal generator}
  \label{fig:qpsk_generator}
\end{figure}

\begin{figure}[H]
  \centering
  \includegraphics[width=\linewidth]{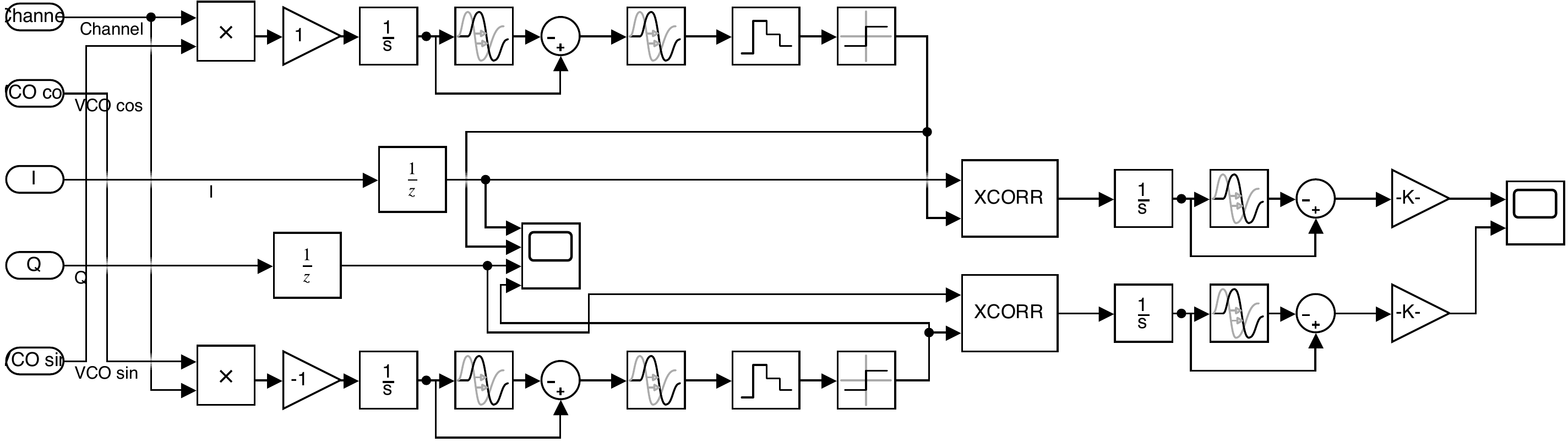}
  \caption{QPSK signal demodulator}
  \label{fig:qpsk_demodulator}
\end{figure}

\begin{figure}[H]
  \centering
  \includegraphics[width=\linewidth]{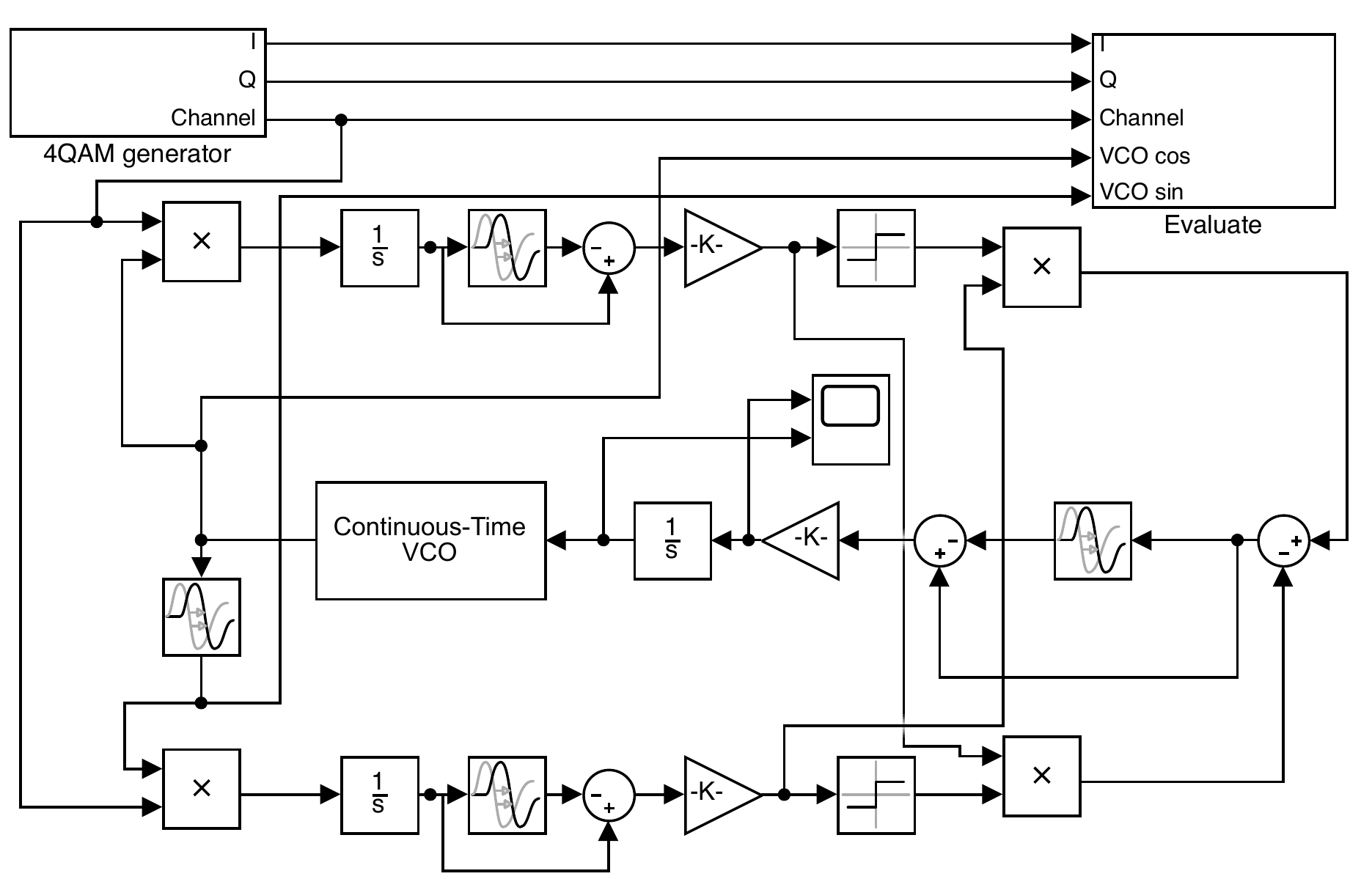}
  \caption{Classic QPSK Costas loop}
  \label{fig:qpsk_classic}
\end{figure}

\begin{figure}[!htp]
  \centering
  \includegraphics[width=\linewidth]{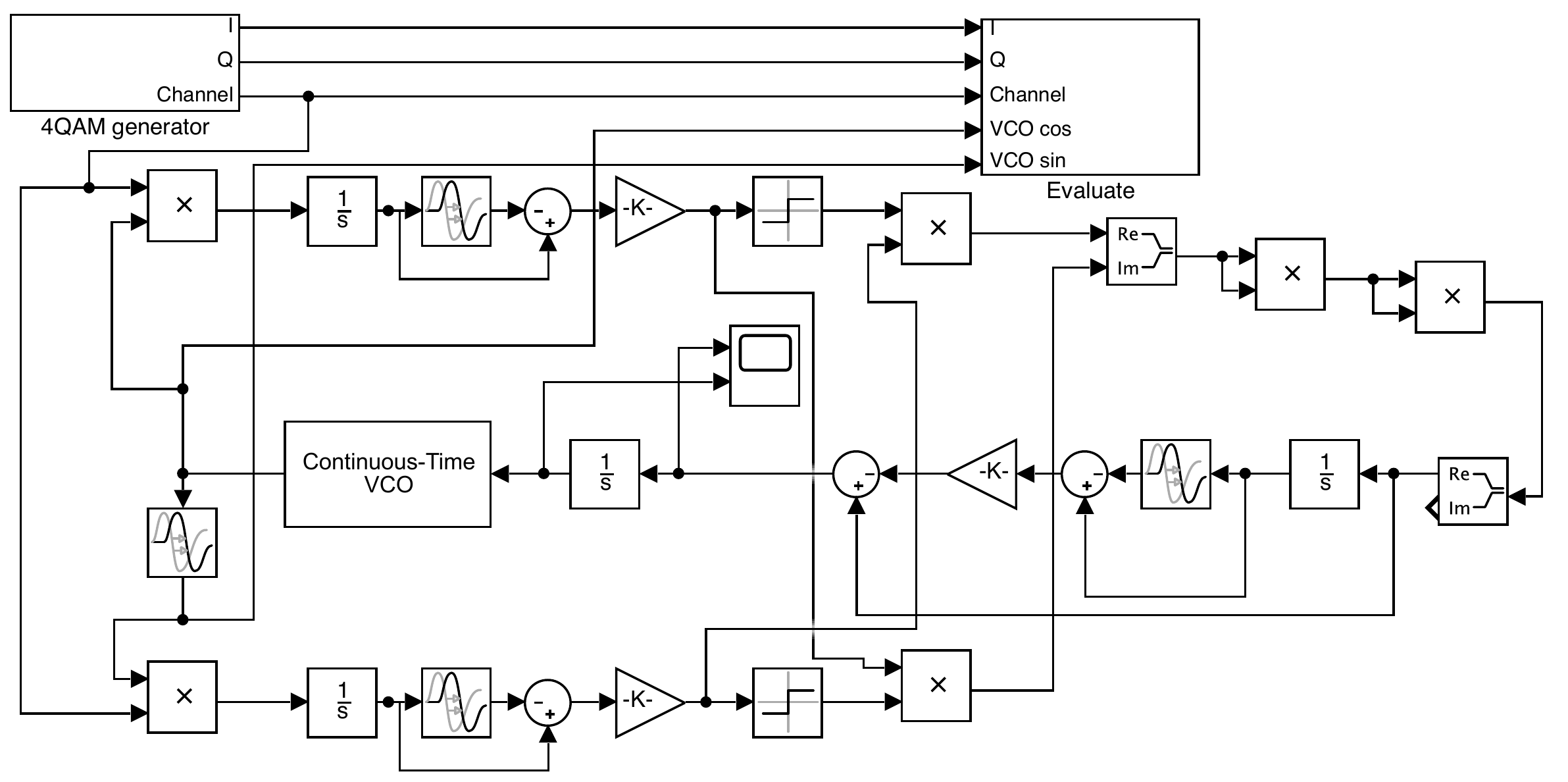}
  \caption{4th power QPSK Costas loop}
  \label{fig:qpsk_4thpower}
\end{figure}

Implementation of classical, 4th power and Folding QPSK Costas modifications are on Fig.~\ref{fig:qpsk_classic}, Fig.~\ref{fig:qpsk_classic}, and Fig.~\ref{fig:qpsk_4thpower}
(see \cite{Ladvanszky-2017}).
Implementations of modulator and demodulator are on Fig~\ref{fig:qpsk_generator} and Fig.~\ref{fig:qpsk_demodulator} correspondingly.
The symbol error rate analysis is presented on Fig.~\ref{ser-snr}.
QPSK Signal generator is composed of Simulink standard library. Random integer block generate digital data, which is then converted to QPSK signal by ``Rectangular 4-QAM'' block, and low-pass filtered by ``Low-pass RF filter''.
Low-pass filtered signal is then phase-shifted by $180$ degrees and
multiplied by carrier (separately Re and Im parts of complex signal).
Small noise is added to resulting signal (``Band-limited white noise'' block).
\begin{figure}[H]
  \centering
  \includegraphics[width=\linewidth]{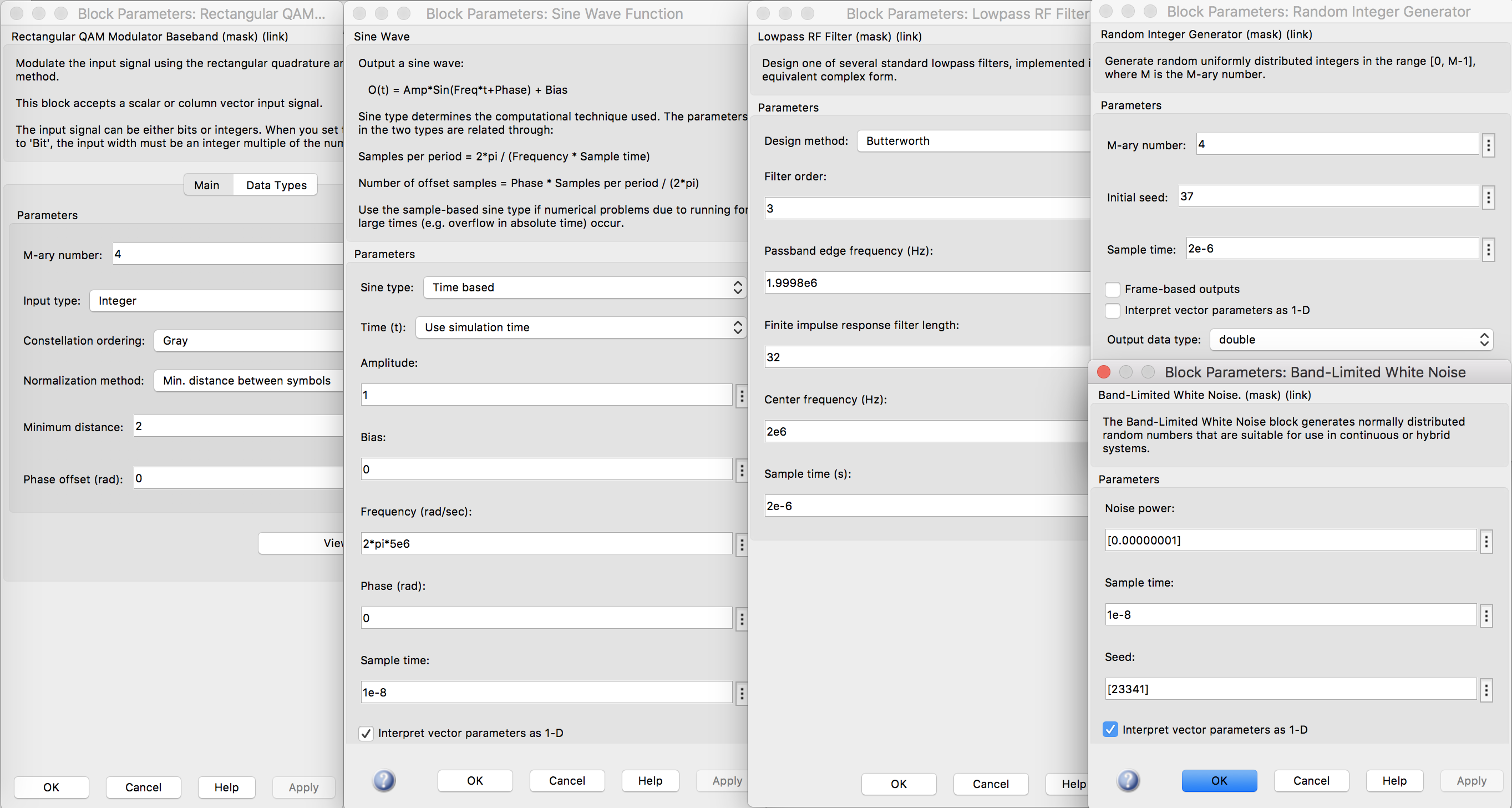}
  \caption{QPSK signal generator: parameters}
  \label{fig:qpsk_generator_params}
\end{figure}
Corresponding block-parameters are shown in Fig.~\ref{fig:qpsk_generator_params}.

Structure of signal demodulator is not the purpose of this study,
therefore here we describe only some example without detailed description.
Signal demodulator multiplies input QPSK signal by the
$\sin$ and $\cos$ carriers, recovered by Costas loop.
Resulting signal is flipped (gain block ``$-1$'') and
 integrated ($\frac{1}{s}$ block).
 After that simple filter is used to obtain moments of
 switching between $0$ and $1$ (delay and difference blocks).
 Next, corresponding signals are converted to digital domain
 by ``Zero-Order Hold'' and ``Sign'' blocks.
 Finally, obtained digital signals are compared
 with original $I$ and $Q$ digital data signals
 (``XCORR'' correlation block, digital filter, and gain blocks).

Consider now classic QPSK Costas loop (fig.~\ref{fig:costas_after_sync}) in Matlab Simulink in Fig.~\ref{fig:qpsk_classic}.
Blocks ``4QAM generator'' and ``Evaluate'' are described above.
Here VCO has single output, and 90 phase-shifter is realized by delay block.
Multiplier block is self-explanatory.
Low-pass filters consist of integration and delay blocks.

Consider 4th power variation of Costas loop in Matlab Simulink (Fig.~\ref{fig:qpsk_4thpower}).
This model is constructed similar to classic QPSK Costas loop.
The 4th power blocks are realized by two multipliers.

\begin{figure}[H]
  \centering
  \includegraphics[width=\linewidth]{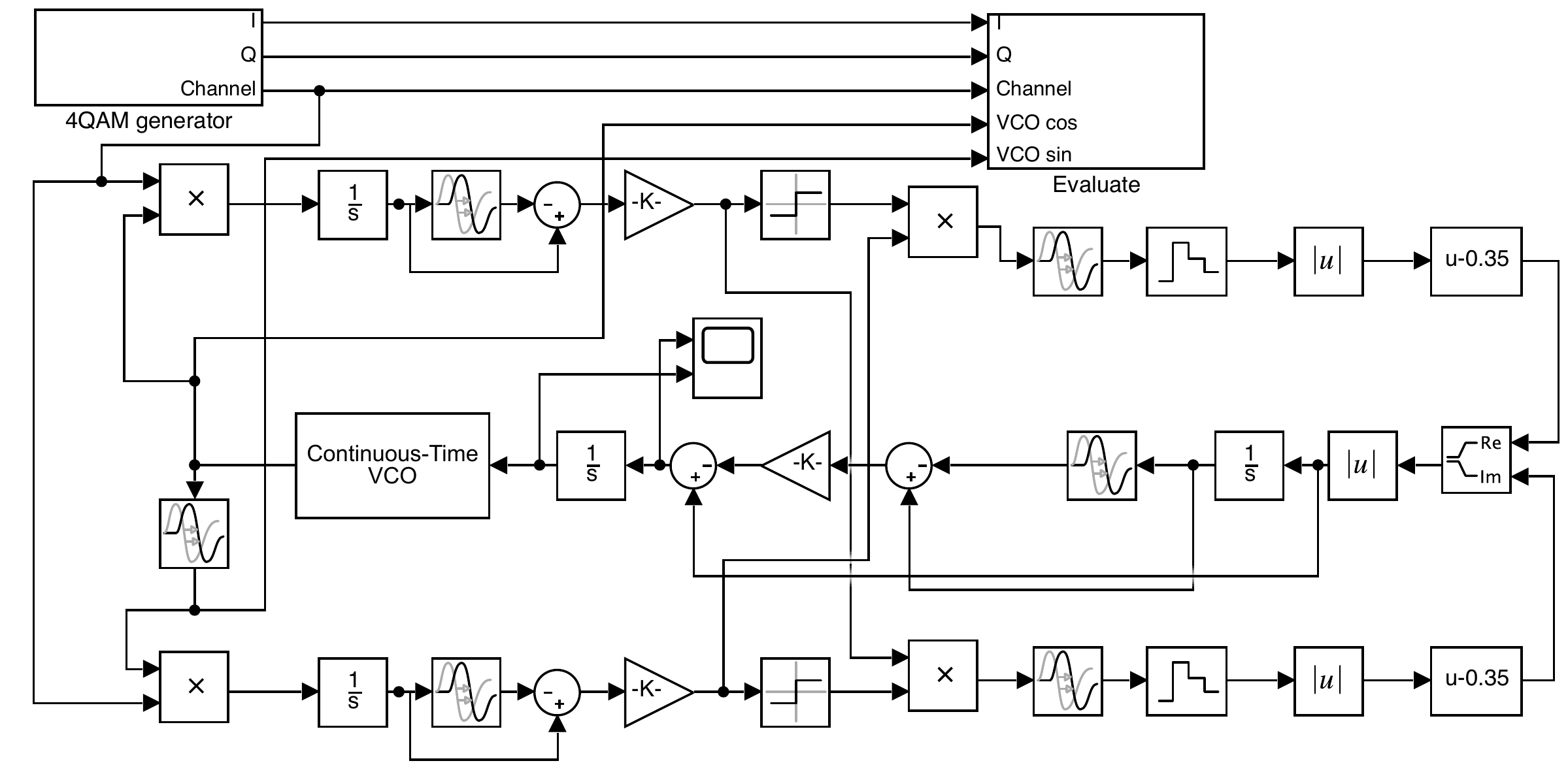}
  \caption{Folding QPSK Costas loop}
  \label{fig:qpsk_folding}
\end{figure}

Finally, Matlab Simulink model for folding QPSK Costas loop
is shown in Fig.~\ref{fig:qpsk_folding}.
Here $|\cdot|$ block is standard abs block with enabled zero-crossing detection
and bias block centers its input.

\section{Conclusions}
\begin{figure}[H]
  \centering
  \includegraphics[width=\linewidth]{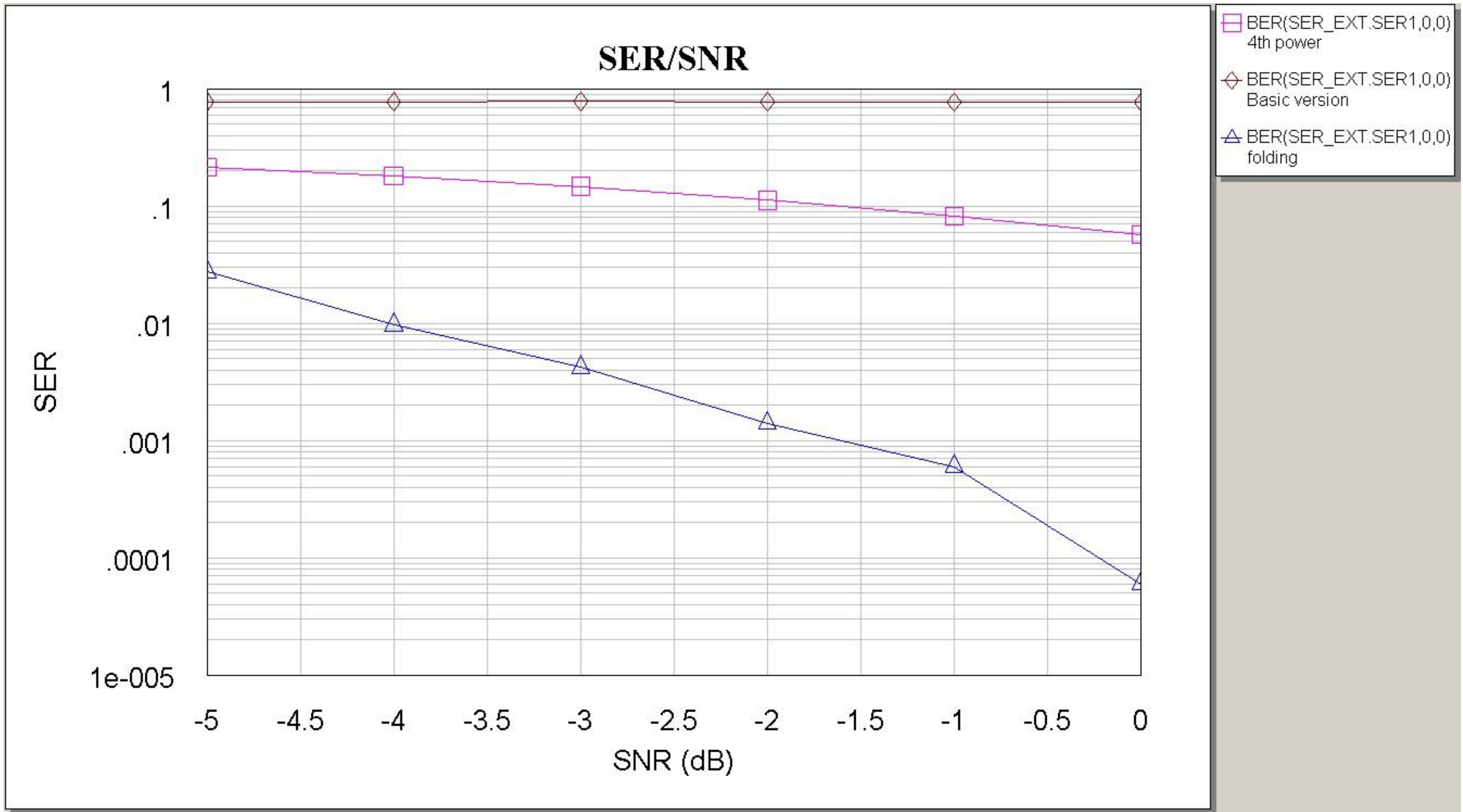}
  \caption{Symbol error rate (SER) comparison in NI AWR Design Environment}
  \label{ser-snr}
\end{figure}

For conclusion consider comparative graph of signal-to-noise analysis
of all three Costas loops (see Fig.~\ref{ser-snr}).

Three modifications of 4QAM(QPSK) Costas loops are considered.
Simplified version of folding variant is presented for the first time.
Hold-in, pull-in, and lock-in ranges for PI loop filters are computed.
It is shown that the folding modification works better in the presence of noise.
More sophisticated noise analysis is given in \cite{SidorkinaSSS-2016}.

More rigorous mathematical models of the QPSK Costas loop circuits
consist of 4th order non-linear differential equations
and corresponding lock-in ranges can be estimated numerically
similar to \cite{LeonovKYY-2015-TCAS,KuznetsovKLYY-2017}.

\section*{Acknowledgements} \label{sec:acknowledgement}
The work is supported by the Russian Science Foundation
project 14-21-00041 (sections 2--4)
and the Leading Scientific Schools of Russia project NSh-2858.2018.1  (section 1).

\end{document}